\documentstyle[amssymb,multicol,amsmath]{article}

\newtheorem{Lemma}{Lemma}

\newcommand{\bp}{\begin{Lemma}}
\newcommand{\ep}{\end{Lemma}}
\newcommand{\bi}[1]{\vspace{-3mm} \bibitem{#1}}
\begin{document}
%%%%%%%%%%%%%%%%%%%%%%%%%%%%%%%%%%%%%%%%%%%%%%%%%%%%%%%%%%%%%%%%%%%%%%
%%%%%%%%%%%%%%%%%%%%%%%%%%%%%%%%%%%%%%%%%%%%%%%%%%%%%%%%%%%%%%%%%%%%%%
%%%%%%%%%%%%%%%%%%%%%%%%%%%%%%%%%%%%%%%%%%%%%%%%%%%%%%%%%%%%%%%%%%%%%%

%%\title{Quantum Computation by Quantum Operations on Mixed States}
\begin{center}
{\Large \bf Quantum Computations by Quantum Operations on 
Mixed States}
\vskip 5 mm
%%\author{Vasily E. Tarasov}
{\Large \bf Vasily E. Tarasov } \\

%\vskip 5 mm
%%{\it Skobeltsyn Institute of Nuclear Physics,
%%Moscow State University, Moscow 119899, Russia }
{\it Skobeltsyn Institute of Nuclear Physics,
Moscow State University, Moscow 119899, Russia}

%\vskip 5 mm
%%\email{TARASOV@THEORY.SINP.MSU.RU}
{E-mail: TARASOV@THEORY.SINP.MSU.RU}
\end{center}
%%\vskip 21 mm

%             %  but any date may be explicitly specified

%%\vskip 5 mm
\begin{abstract}
{\small Usually models for quantum computations deal with unitary gates
on pure states. In this paper we generalize the usual model.  
We consider a model of quantum computations  
in which the state is an operator of density
matrix and the gates are quantum
operations, not necessarily unitary. A mixed state (operator of
density matrix) of  n two-level quantum systems 
is considered as an element of $4^{n}$-dimensional operator Hilbert space.
Unitary quantum gates and nonunitary quantum operations for
n-qubit system
are considered as generalized quantum gates acting on mixed state.
In this paper we study universality for quantum computations by 
quantum operations on mixed states.
}
\end{abstract}

%%\vskip 11 mm
PACS {3.67.Lx}
% 03.65.-w fundamental quantum mechanics
% 03.65.Bz Foundations, theory of measurement, miscellaneous
% 03.67.-a quantum information
% 03.67.Lx quantum computation
% 89.70.+c Information science

%%\vskip 3 mm
%%\keywords{Quantum computations, quantum gates, mixed states, quantum operations}

%%%%%%%%%%%%%%%%%%%%%%%%%%%%%%%%%%%%%%%%%%%%%%%%%%%%%%%%%%%%%%%%%%%%%%%%%%

%%\vskip 44mm

%%\newpage
%%%%%\twocolumn

%%%%%%%%%%%%%%%%%%%%%%%%%%%%%%%%%%%%%%%%%%%%%%%%%%%%%%%%%%%%%%%%%%%%%%%%%%
%%%%%%%%%%%%%%%%%%%%%%%%%%%%%%%%%%%%%%%%%%%%%%%%%%%%%%%%%%%%%%%%%%%%%%%%%%
%%%%%%%%%%%%%%%%%%%%%%%%%%%%%%%%%%%%%%%%%%%%%%%%%%%%%%%%%%%%%%%%%%%%%%%%%%

\begin{multicols}{2}

%%%%%%%%%%%%%%%%%%%%%%%%%%%%%%%%%%%%%%%%%%%%%%%%%%%%%%%%%%%%%%%%%%%%%%%%%%
%%%%%%%%%%%%%%%%%%%%%%%%%%%%%%%%%%%%%%%%%%%%%%%%%%%%%%%%%%%%%%%%%%%%%%%%%%

\section{Introduction}

%%%%%%%%%%%%%%%%%%%%%%%%%%%%%%%%%%%%%%%%%%%%%%%%%%%%%%%%%%%%%%%%%%%%%%%%%

Usual models for quantum computations deal only with unitary gates
on pure states. In these models it is difficult or impossible
to deal formally with measurements, dissipation, decoherence and noise.
Understanding dynamics of mixed states is important for studying
quantum noise processes \cite{Gar,Schu,SN}, quantum error
correction \cite{Sh,St2,BDSW}, decoherence effects
\cite{Unr,MPP,BBSS,LCW,Zan} in quantum computations
and to perform simulations of open quantum systems
\cite{Ter,TDV,Bac,LV,AbL,Zal}.
It turns out, that the restriction to pure states and unitary gates
is unnecessary \cite{AKN,Tarpr}.

In this paper we generalize the usual model of quantum computations
to a model in which the state is a density matrix operator
and the gates are general quantum operations,
not necessarily unitary.
Pure state of n two-level quantum systems is an element
of $2^{n}$-dimensional Hilbert space. Usually the gates of this
model are unitary operators act on a such state.
In general case, mixed state (operator of density matrix)
of  n two-level quantum systems is an element of
$4^{n}$-dimensional operator Hilbert space.
The gates for mixed states are general quantum
operations which act on general mixed states. 
Unitary gates and quantum operations for quantum
two-valued logic computations are considered as four-valued logic
gates of new model. 
The space of linear operators acting on a
$N=2^{n}$-dimensional Hilbert space
is a $N^{2}=4^{n}$-dimensional operator Hilbert space.
%%It leads to quantum computer model with 2-valued logic.
The mixed state of $n$ two-level
quantum system is an element of $4^{n}$-dimensional
operator Hilbert space.
It leads to 4-valued logic model for quantum computations 
with mixed states.
In the paper we consider universality for
general quantum gates acting on mixed states.
The condition of completely positivity leads to difficult
inequalities for gate matrix elements \cite{Choi,FA,KR,RSW}. In
order to satisfy condition of completely positivity we use the
following representation. 
Any linear completely positive quantum operation can be represented by
\[ \hat{\cal E}=\sum^{m}_{j=1} 
\hat L_{A_j} \hat R_{A^{\dagger}_{j}}: \quad 
\hat{\cal E}(\rho)=\sum^{m}_{j=1} 
A_j \rho A^{\dagger}_{j}, \]
where $\hat L_A B=AB$ and $\hat R_A B=BA$.
To find the universal set of completely positive 
(linear or nonlinear) gates $\hat{\cal E}$ 
we consider the universal set of  $\hat L_{A_j}$
and $\hat R_{A^{\dagger}_j}$.
A two-qubit gate $\hat{\cal E}$ is called primitive if
$\hat{\cal E}$ maps tensor product of single ququats
to tensor product of single qubits.
The gate $ \hat{\cal E}$ is called imprimitive if
$\hat{\cal E}$ is not primitive.
We prove that almost every pseudo-gate that operates
on two or more ququats is universal pseudo-gate.
The set of all single ququat pseudo-gates and
any imprimitive two-ququats pseudo-gate
are universal set of pseudo-gates.

In Section 2, we introduce generalized computational basis and
generalized computational states for $4^{n}$-dimensional operator
Hilbert space. In the Section 3, we study some
properties of general quantum gates. General
quantum operations are considered as
generalized quantum gates. 
In the Section 4, we consider a
universal set of quantum 4-valued logic gates. 
In the Section 5, unitary 2-valued logic gates 
are considered as generalized quantum gates.
We realize classical
4-valued logic gates by quantum gates. 
In Appendix 1, the physical and mathematical background
(pure and mixed states, operator Hilbert space and superoperators) 
are considered.
In Appendix 2, we introduce a four-valued
classical logic formalism.

%%%%%%%%%%%%%%%%%%%%%%%%%%%%%%%%%%%%%%%%%%%%%%%%%%%%%%%%%%%%%%%%%%%%%%%%%
%%%%%%%%%%%%%%%%%%%%%%%%%%%%%%%%%%%%%%%%%%%%%%%%%%%%%%%%%%%%%%%%%%%%%%%%%

\section{Computational basis for mixed states}

%%%%%%%%%%%%%%%%%%%%%%%%%%%%%%%%%%%%%%%%%%%%%%%%%%%%%%%%%%%%%%%%%%%%%%%%%

%%\subsection{Mixed states}
%%{\bf 2.2. Mixed states}\\

In general, a quantum system is not in a pure state.
Landau and von Neumann
introduced a mixed state and a density matrix into quantum theory.
A density matrix is a Hermitian ($\rho^{\dagger}=\rho$), positive
($\rho >0$) operator on ${\cal H}^{(n)}$ with trace $Tr \rho =1$.
Pure states can be characterized as orthogonal projections of unit
trace: $\rho^{2}=\rho$, $\rho^{\dagger}=\rho$, $Tr \rho =1$. A
pure state is represented by the operator
$\rho=|\Psi><\Psi|$.

One can represent an arbitrary density matrix operator $\rho(t)$
for $n$-qubits (n two-level quantum systems) in terms 
of tensor products of Pauli matrices $\sigma_{\mu}$:
\begin{equation}
\label{rhosigma} \rho(t)=\frac{1}{2^{n}} \sum_{\mu_1 ... \mu_n}
P_{\mu_1 ... \mu_n}(t)
\sigma_{\mu_1} \otimes ... \otimes \sigma_{\mu_n} \ . \end{equation}
where each $\mu_{i} \in \{0,1,2,3\}$, $i=1,...,n$ and $\sigma_0=I$.
If $\mu_i=1,2,3$, then $\sigma_{\mu_i}$ are Pauli matrices.

%%$$ \label{sigma}
%%\sigma_{1}=\left(
%%\begin{array}{cc}
%%0&1\\
%%1&0\\
%%\end{array}
%%\right), \ \ \
%%%%\quad
%%\sigma_{2}=\left(
%%\begin{array}{cc}
%%0&-i\\
%%i&0\\
%%\end{array}
%%\right), $$
%%%%\quad
%%$$\sigma_{3}=\left(
%%\begin{array}{cc}
%%1&0\\
%%0&-1\\
%%\end{array}
%%\right), \ \ \
%%%%\quad
%%\sigma_{0}=I=\left(
%%\begin{array}{cc}
%%1&0\\
%%0&1\\
%%\end{array}
%%\right).
%%$$
The real expansion coefficients $P_{\mu_1 ... \mu_n}(t)$ are given by
\[ P_{\mu_1 ... \mu_n}(t)=Tr( \sigma_{\mu_1} \otimes ...
\otimes \sigma_{\mu_n} \rho(t)). \] Normalization ($Tr \rho=1$)
requires that $P_{0...0}(t)=1$. Since the eigenvalues of the Pauli
matrices are $\pm 1$, the expansion coefficients satisfy
$|P_{\mu_1...\mu_n}(t)|\le 1$. Let us rewrite (\ref{rhosigma}) in
the form:
\begin{equation} \label{rho2} \rho(t)=
\frac{1}{2^n}\sum^{N-1}_{\mu=0} \sigma_{\mu} P_{\mu}(t), \end{equation}
where $\sigma_{\mu}=\sigma_{\mu_1} \otimes ... \otimes \sigma_{\mu_n}$,
$\mu=(\mu_1...\mu_n)$ and $N=4^{n}$.

Let us introduce generalized computational basis and generalized
computational  states for $4^{n}$-dimensional operator Hilbert space.
For the concept of operator Hilbert space and superoperators see
Aappendix 1 and \cite{Cra}-\cite{kn2}. 
Pauli matrices can be considered as a basis of
operator Hilbert space (see Appendix 1).

We can rewrite formulas (\ref{rho2}) using the complete
operator basis $|\sigma_{\mu})$ in operator Hilbert space $\overline{\cal
H}^{(n)}$:
\[ |\rho(t))=\frac{1}{2^n}\sum^{N-1}_{\mu=0}
|\sigma_{\mu}) P_{\mu}(t), \]
where $P_{\mu}(t)=(\sigma_{\mu}|\rho(t))$.
The basis
$|\sigma_{\mu})$ is orthogonal, but is not orthonornal.
Let us define the orthonormal basis $|\mu)$ of operator Hilbert space
$\overline{\cal H}^{(n)}$.
The basis for $\overline{\cal H}^{(n)}$ consists of the
$N^2=4^n$ orthonormal basis elements denoted by $|\mu)$.

\noindent {\bf Definition}
{\it A basis of operator Hilbert space $\overline{\cal H}^{(n)}$ is defined by
\[ |\mu)=|\mu_1...\mu_n)=\frac{1}{\sqrt{2^{n}}}|\sigma_{\mu})=
\frac{1}{\sqrt{2^{n}}}|\sigma_{\mu_1}
\otimes  ... \otimes \sigma_{\mu_n}), \]
where each $\mu_i \in \{0,1,2,3\}$, $N=4^n$ and
\[ (\mu|\mu')=\delta_{\mu \mu'} \ , \quad
\sum^{N-1}_{\mu=0} |\mu)(\mu|=\hat I, \]
is called a {\bf generalized computational basis}.
Here $\mu$ is  4-valued representation of
$\mu=\mu_1 4^{n-1}+...+\mu_{n-1}4+\mu_n$.}

\noindent
{\bf Example.}
In general case, one-qubit mixed state $\rho(t)$ is
\[ |\rho)=|0)\frac{1}{\sqrt{2}}+|1)\rho_{1}+
|2)\rho_{2}+|3) \rho_{3}, \]
where {\it four} orthonormal basis elements are
$|\mu)=(1/\sqrt{2}) |\sigma_\mu)$, $(\mu=0,1,2,3)$.

The usual computational basis $\{|k>\}$ is not a basis of general
state $\rho(t)$ which has a time dependence. In general case, a
pure state evolves to mixed state.

Pure state of n two-level quantum systems is an element
of $2^{n}$-dimensional functional Hilbert space ${\cal H}^{(n)}$.
It leads to model of quantum computations with 2-valued logic.
{\it In general case, the mixed state $\rho(t)$ of $n$ two-level
quantum system is an element of $4^{n}$-dimensional
operator Hilbert space $\overline{\cal H}^{(n)}$.
It leads to 4-valued logic model for quantum computations.}

The state $|\rho(t))$ at any point time is a superposition
of basis elements
\[ |\rho(t))=\sum^{N-1}_{\mu=0} |\mu)\rho_{\mu}(t), \]
where $\rho_{\mu}(t)$ are real numbers (functions)
\[ \rho_{\mu}(t)=(\mu|\rho(t))=\frac{1}{\sqrt{2^n}}
Tr(\sigma_{\mu} \rho(t)). \]
Note that $\rho_0(t)=(0|\rho(t))=1/\sqrt{2^{n}}Tr\rho(t)=1/\sqrt{2^{n}}$
for all cases.

Generalized computational
basis elements $|\mu)$ are not quantum states for
$\mu\not=0$. It follows from normalized condition
$(0|\rho(t))=1/\sqrt{2}$.
Let us define simple computational quantum states.

\noindent {\bf Definition}
{\it A quantum states in operator Hilbert space defined by
\[ |\mu]=\frac{1}{\sqrt{2^{n}}}\Bigl(|0)+
|\mu)(1-\delta_{\mu 0}) \Bigr). \]
are called {\bf generalized computational states}.}

Note that all states $|\mu]$, where $\mu \not=0$, are pure states,
since $[\mu|\mu]=1$. The state $|0]$ is maximally mixed state.
The states $|\mu]$ are elements of operator Hilbert space
$\overline{\cal H}^{(n)}$.

A state in a 4-dimensional Hilbert space can be called
ququat (quantum quaternary digit).
Usually ququat is considered as 4-level quantum system. We
consider ququat as general state (density matrix operator) 
in a 4-dimensional operator Hilbert space.

\noindent {\bf Definition}
{\it A quantum state in 4-dimensional operator Hilbert space
$\overline{\cal H}^{(1)}$ associated with single qubit of 
${\cal H}^{(1)}={\cal H}_2$ is called {\bf single ququat}.
A quantum state in $4^n$-dimensional operator Hilbert space
$\overline{\cal H}^{(n)}$ associated with n-qubits system
is called {\bf n-ququats}.}

In this case quantum operations and unitary 2-valued logic quantum
gates can be considered as quantum 4-valued logic gates
acting on n-ququats.

\noindent
{\bf Example.}
For the single ququat the states $|\mu]$ are
\[ |0]=\frac{1}{\sqrt{2}}|0)  \ , \quad
|k]=\frac{1}{\sqrt{2}}\Bigl(|0)+|k)\Bigr). \]

It is convenient to use matrices for quantum states. In 
matrix representation the single ququat computational basis
$|\mu)$ and computational states $|\mu]$ can be represented by
column \cite{Tarpr}.

We can use the other matrix representation for the states
$|\rho]$ which has no the coefficient $1/\sqrt{2^{n}}$.
The single qubit generalized computational
states $|\mu]$ can be represented by column of $1,P_1,P_2,P_3$.
A general single ququat quantum state
$|\rho]$ is a superposition
of generalized computational states
\[ |\rho]=|0](1-P_1-P_2-P_3)+|1] P_1+|2]P_2+|3] P_3. \]

%%%%%%%%%%%%%%%%%%%%%%%%%%%%%%%%%%%%%%%%%%%%%%%%%%%%%%%%%%%%%%%%%%%%%%%%

\section{Quantum operations as quantum gates}

%%%%%%%%%%%%%%%%%%%%%%%%%%%%%%%%%%%%%%%%%%%%%%%%%%%%%%%%%%%%%%%%%%%%%%%%

In this section we consider some properties of quantum operations
as four-valued logic gates. 

%%%%%%%
 
Unitary evolution is not the most general type
of state change possible for quantum systems.
A most general state change of a quantum system
is a positive trace-preserving map which
is called a quantum operation or superoperator.
For the concept of quantum operations see \cite{Kr3,Kr4,Schu}.
In the formalism of quantum operations the final 
state $\rho^{\prime}$ is related to the initial state $\rho$
by a map
\begin{equation} \label{Et} \rho \ \rightarrow \ \rho^{\prime}=
\frac{{\cal E}(\rho)}{Tr({\cal E}(\rho) )} \ . \end{equation} The
trace in the denominator is induced in order to preserve the trace
condition, $Tr(\rho^{\prime})=1$. In general case, this
denominator leads to the map is nonlinear, where the map 
${\cal E}$ is a linear positive map.

The quantum operation ${\cal E}$
usually considered as a completely positive map.
The most general form for completely positive quantum
operation ${\cal E}$ is
\[ {\cal E}(\rho)=\sum^{m}_{j=1} A_j \rho A^{\dagger}_j. \]
By definition, $Tr({\cal E}(\rho))$ is the probability that the process
represented by ${\cal E}$ occurs, when $\rho$ is the initial state.
The probability never exceed 1. The quantum operation ${\cal E}$
is trace-decreasing, i.e. $Tr({\cal E}(\rho)) \le 1$ for all
density matrix operators $\rho$. This condition can be expressed
as an operator inequality for $A_{j}$.
The operators $A_j$ must satisfy
\[ \sum^{m}_{j=1} A^{\dagger}_j A_{j} \le I. \]
%%\[ \hat{\cal E}^{\dagger}(I)=\sum^{m}_{j=1} A^{\dagger}_j A_{j} \le I. \]
The normalized post-dynamics system state is defined by (\ref{Et}).
The map (\ref{Et}) is nonlinear trace-preserving map.
If the linear quantum operation ${\cal E}$ is trace-preserving
$Tr({\cal E}(\rho))=1$, then
\[ \sum^{m}_{j=1} A^{\dagger}_j A_{j}=I. \]
Notice that a trace-preserving quantum operation
${\cal E}(\rho)=A\rho A^{\dagger}$ must be a unitary
transformation ($A^{\dagger}A=AA^{\dagger}=I$).

Quantum operations can be considered as generalized quantum gates
act on mixed states.
Let us define a generalized quantum gates.

\noindent {\bf Definition}
{\it Quantum (four-valued logic) gate is
a superoperator $\hat{\cal E}$ on operator Hilbert space
$\overline{\cal H}^{(n)}$ which maps a density
matrix operator $|\rho)$ of $n$-ququats to a density matrix operator
$|\rho')$ of $n$-ququats.}

A generalized quantum gate is a superoperator $\hat{\cal E}$ which
maps density matrix operator $|\rho)$ to density matrix operator
$|\rho^{\prime})$.
If $\rho$ is operator of density matrix, then $\hat{\cal E}(\rho)$
should also be a density matrix operator.
Any density matrix operator is self-adjoint ($\rho^{\dagger}(t)=\rho(t)$),
positive ($\rho(t)>0$) operator with unit trace ($ Tr\rho(t)=1$).
Therefore we have some requirements for superoperator $\hat{\cal E}$.

The requirements for a superoperator $\hat{\cal E}$
to be a generalized quantum gate are as follows: \\
1. The superoperator $\hat{\cal E}$ is {\it real} superoperator, i.e.
$\Bigl(\hat{\cal E}(A)\Bigr)^{\dagger}=\hat{\cal E}(A^{\dagger})$
for all $A$ or $\Bigl(\hat{\cal E}(\rho)\Bigr)^{\dagger}=\hat{\cal E}(\rho)$.
Real superoperator $\hat{\cal E}$ maps self-adjoint operator
$\rho$ into self-adjoint operator $\hat{\cal E}(\rho)$:
$ (\hat{\cal E}(\rho))^{\dagger}=\hat{\cal E}(\rho)$.
\\
2.1. The gate $\hat{\cal E}$ is a {\it positive} superoperator,
i.e. $\hat{\cal E}$ maps positive operators to positive operators:
\ $\hat{\cal E}(A^{2}) >0$ for all $A\not=0$ or $\hat{\cal E}(\rho)
\ge 0$.\\ 2.2. We have to assume the superoperator $\hat{\cal E}$
to be not merely positive but completely positive. The
superoperator $\hat{\cal E}$ is {\it completely positive} map of
operator Hilbert space, i.e. the positivity is remained if we extend the
operator Hilbert space $\overline{\cal H}^{(n)}$ by adding more qubits.
That is, the superoperator $\hat{\cal E} \otimes \hat I^{(m)}$
must be positive, where $\hat I^{(m)}$ is the identity
superoperator on some operator Hilbert space $\overline{\cal H}^{(m)}$.\\
3. The superoperator $\hat{\cal E}$ is {\it trace-preserving} map, i.e.
\[ (I|\hat{\cal E}|\rho)=(\hat{\cal E}^{\dagger}(I)|\rho)=1 \quad
or \quad \hat{\cal E}^{\dagger}(I)=I. \]
3.1. The superoperator $\hat{\cal E}$ is a {\it linear} map
of density matrix operators.
Any linear completely positive superoperator can be represented by
\begin{equation}  \label{elr}
\hat{\cal E}=\sum^{m}_{j=1} \hat L_{A_j} \hat R_{A^{\dagger}_{j}}
\end{equation}
3.2. The restriction to linear gates is unnecessary.
Let us consider $\hat{\cal E}$ is a linear superoperator which is not
trace-preserving. This superoperator is not a quantum gate.
Let $(I|\hat{\cal E}|\rho)=Tr(\hat{\cal E}(\rho))$
is a probability that the process represented
by the superoperator $\hat{\cal E}$ occurs.
Since the probability is nonnegative and
never exceed 1, it follows that the superoperator
$\hat{\cal E}$ is a trace-decreasing superoperator:
\[ 0\le (I|\hat{\cal E}|\rho) \le 1 \quad
or \quad \hat{\cal E}^{\dagger}(I) \le I. \]
In general case, the linear trace-decreasing superoperator is not a
quantum four-valued logic gate, since it can be not trace-preserving.
The generalized quantum gate can be defined as {\it nonlinear
trace-preserving} gate $\hat{\cal N}$ by
\[ \hat{\cal N}|\rho)=
(I|\hat{\cal E}|\rho)^{-1} \hat{\cal E}|\rho) \quad or
\quad \hat{\cal N}(\rho)=
\frac{\hat{\cal E}(\rho)}{Tr(\hat{\cal E}(\rho))}, \]
where $\hat{\cal E}$ is a linear completely positive trace-decreasing
superoperator.

Four-valued logic gates $\hat{\cal E}$
in the matrix representation can be represented by $4^{n}\times 4^{n}$
matrices ${\cal E}_{\mu \nu}$.
In this matrix representation the gate $\hat{\cal E}$ maps
the state $|\rho(t_0))=\sum^{N-1}_{\nu=0}|\nu) \rho_{\nu}(t_0)$
to the state $|\rho(t))=\sum^{N-1}_{\mu}|\mu) \rho_{\mu}(t)$ by
\begin{equation} \label{rEr} 
\rho_{\mu}(t)=\sum^{N-1}_{\nu=0}
{\cal E}_{\mu \nu} \rho_{\nu}(t_0) \ . \end{equation}
where $\rho_{0}(t)=\rho_{0}(t_0)=1/\sqrt{2^{n}}$ and $N=4^n$.
Since $P_{\mu}(t_0)=\sqrt{2^{n}} \rho_{\mu}(t_0)$ and
$P_{\mu}(t)=\sqrt{2^{n}} \rho_{\mu}(t)$,
it follows that relation (\ref{rEr}) for linear gate
$\hat{\cal E}$ is equivalent to
\begin{equation} \label{PEP} P_{\mu}(t)=\sum^{N-1}_{\nu=0}
{\cal E}_{\mu \nu} P_{\nu}(t_0) \ . \end{equation}

%%{\bf Lemma 1.}\\
\bp
{\it In the generalized computational basis $|\mu)$ any linear
two-valued logic quantum operation ${\cal E}$ can be represented as
a  quantum four-valued logic gate $\hat{\cal E}$ defined by
\[ \hat{\cal E}=\sum^{N-1}_{\mu=0}\sum^{N-1}_{\nu=0}
{\cal E}_{\mu \nu} \ |\mu)(\nu|  \ , \]
where $N=4^n$,
\[ {\cal E}_{\mu \nu}=\frac{1}{2^n}
Tr\Bigl(\sigma_{\mu} \hat{\cal E} (\sigma_{\nu}) \Bigr), \]
and  $\sigma_{\mu} =\sigma_{\mu_1} \otimes ...  \otimes \sigma_{\mu_n}$.

Here $N=4^n$, $\mu$ and $\nu$ are 4-valued representation of
\[ \mu=\mu_1 4^{N-1}+...+\mu_{N-1}4+\mu_N , \]
\[ \nu=\nu_1 4^{N-1}+...+\nu_{N-1}4+\nu_N, \]
$\mu_i,\nu_i \in \{0,1,2,3\}$ and
${\cal E}_{\mu \nu}$ are elements of some matrix.
}
\ep

\noindent {\bf Proof.}
The state $\rho(t)$ in the generalized computational basis
$|\mu)$ has the form
\[ |\rho(t))=\sum^{N-1}_{\mu=0} |\mu)\rho_{\mu}(t)  \ , \]
where $N=4^{n}$ and
$\rho_{\mu}(t)=(\mu|\rho(t))$.
The quantum operation ${\cal E}$ defines a four-valued logic gate by
\[ |\rho(t))= \hat{\cal E}_{t}|\rho) =| {\cal E}_{t}(\rho) )=
\sum^{N-1}_{\nu=0} |{\cal E}_{t}(\sigma_{\nu}) )
\frac{1}{\sqrt{2^n}} \rho_{\nu}(t_{0}). \]
Then
\[ (\mu|\rho(t))=\sum^{N-1}_{\nu =0}
(\sigma_{\mu}|{\cal E}_{t}( \sigma_{\nu}) )
\frac{1}{2^n} \rho_{\nu}(t_{0}). \]
Finally, we obtain (\ref{rEr}),
where
\[ {\cal E}_{\mu \nu}=\frac{1}{2^n}
(\sigma_{\mu}|{\cal E}_{t}(\sigma_{\nu}) )
=\frac{1}{2^n}Tr\Bigl(\sigma_{\mu} {\cal E}_{t}(\sigma_{\nu}) \Bigr). \]
This formula defines a relation between quantum
operation ${\cal E}$ and the real $4^{n}\times 4^{n}$ matrix
${\cal E}_{\mu \nu}$ of quantum gate $\hat{\cal E}$.

%{\bf Lemma 2.}\\
\bp
{\it In the generalized computational basis $|\mu)$ the matrix
${\cal E}_{\mu \nu}$ of general quantum four-valued
logic gate (\ref{elr}) is real
${\cal E}^{*}_{\mu \nu}={\cal E}_{\mu \nu}$. }
\ep

\noindent {\bf Proof.}
\[ {\cal E}_{\mu \nu}=\frac{1}{2^n}\sum^{m}_{j=1}
Tr\Bigl(\sigma_{\mu} A_{j} \sigma_{\nu} A^{\dagger}_{j} \Bigr)=
\frac{1}{2^n} \sum^{m}_{j=1}
(A^{\dagger}_j  \sigma_{\mu}| \sigma_{\nu}A^{\dagger}_j ). \]
\[ {\cal E}^{*}_{\mu \nu}=\frac{1}{2^n} \sum^{m}_{j=1}
(A^{\dagger}_j \sigma_{\mu}| \sigma_{\nu} A^{\dagger}_j)^{*}=
\frac{1}{2^n} \sum^{m}_{j=1}
( \sigma_{\nu} A^{\dagger}_j|A^{\dagger}_j  \sigma_{\mu} )=  \]
\[ =\frac{1}{2^n} \sum^{m}_{j=1}
Tr\Bigl(A_j \sigma_{\nu} A^{\dagger}_{j} \sigma_{\mu} \Bigr)=
\frac{1}{2^n} \sum^{m}_{j=1}
Tr\Bigl( \sigma_{\mu} A_{j} \sigma_{\nu} A^{\dagger}_j \Bigr)=
{\cal E}_{\mu \nu}. \]

%%{\bf Lemma 3.}\\
\bp
{\it Any real matrix ${\cal E}_{\mu \nu}$ associated with
linear (trace-preserving) quantum gates (\ref{elr}) has
${\cal E}_{0 \nu}=\delta_{0 \nu}$.}
\ep

\noindent {\bf Proof.}
\[ {\cal E}_{0 \nu}=\frac{1}{2^{n}}
Tr\Bigl(\sigma_{0}{\cal E}(\sigma_{\nu}) \Bigr)=
\frac{1}{2^{n}} Tr\Bigl({\cal E}(\sigma_{\nu}) \Bigr)= \]
\[ =\frac{1}{2^{n}}
Tr\Bigl(\sum^{m}_{j=1} A_j \sigma_{\nu}A^{\dagger}_j \Bigr)=
\frac{1}{2^{n}} Tr\Bigl((\sum^{m}_{j=1}
A^{\dagger}_j A_j )\sigma_{\nu} \Bigr)=\]
\[=\frac{1}{2^{n}} Tr \sigma_{\nu}=\delta_{0\nu}. \]
Completely positive condition leads to some inequalities
\cite{KR,RSW,Choi} for matrix elements ${\cal E}_{\mu \nu}$.

Let us consider the n-ququats linear quantum gate
\begin{equation} \label{LGE}
\hat{\cal E}=|0)(0|+\sum^{N-1}_{\mu=1} T_{\mu} |\mu)(0|+
\sum^{N-1}_{\mu=1} \sum^{N-1}_{\nu=1} R_{\mu \nu} |\mu)(\nu|,
\end{equation}
where $N=4^{n}$. 
In general case, linear quantum 4-value logic gate acts on
$|0)$ by
\[ \hat{\cal E}|0)=|0)+\sum^{N-1}_{k=1}T_{k}|k). \]
If all $T_{k}$, where $k=1,...,N-1$ is equal to zero,
then $\hat{\cal E}|0)=|0)$. The linear quantum gates with $T=0$
conserve the maximally mixed state $|0]$ invariant.

\noindent {\bf Definition} {\it A quantum four-valued logic gate
$\hat{\cal E}$ is called unital gate or gate with $T=0$ if
maximally mixed state $|0]$ is invariant under the action of this
gate: $\hat{\cal E}|0]=|0]$.}

The matrix ${\cal E}_{\mu \nu}$ of linear trace-preserving
n-ququats gate $\hat{\cal E}$ is an element of group 
$TGL(4^n-1,\mathbb{R})$
which is a semidirect product of general linear group
$GL(4^n-1,\mathbb{R})$ and translation group $T(4^n-1,\mathbb{R})$.
This proposition follows from Lemma 3.
Any element (gate matrix ${\cal E}_{\mu \nu}$) of group 
$TGL(4^n-1,\mathbb{R})$
can be represented by
$${\cal E}(T,R)=\left(
\begin{array}{cc}
1&0\\
T&R
\end{array}
\right),$$
where $T$ is a column with $4^n-1$ elements,
$0$ is a line with $4^n-1$ zero elements, and
$R$ is a real $(4^n-1)\times (4^n-1)$ matrix
$R \in GL(4^n-1,\mathbb{R})$.
If $R$ is orthogonal $(4^{n}-1)\times(4^{n}-1)$ matrix ($R^TR=I$),
then we have motion group \cite{Vil}.
The group multiplication of elements 
${\cal E}(T,R)$ and ${\cal E}(T',R')$
is defined by
\[ {\cal E}(T,R){\cal E}(T',R')={\cal E}(T+RT',RR'). \]
In particular, we have
\[ {\cal E}(T,R)={\cal E}(T,I){\cal E}(0,R) \ , \quad
{\cal E}(T,R)={\cal E}(0,R){\cal E}(R^{-1}T,I). \]
where $I$ is unit $(4^n-1)\times (4^n-1)$ matrix.

%%%%%%%%%%%%%

Let us consider the n-ququats linear gate (\ref{LGE}).  
The gate matrix ${\cal E}(T,R)$ is an element of Lie
group $TGL(N-1,\mathbb{R})$, here $N=4^{n}$. The  matrix $R$
is an element of Lie group $GL(N-1,\mathbb{R})$.

\noindent {\bf Theorem 1.} (Singular Valued Decomposition for Matrix)\\
{\it Any real  matrix $R$ can be written in the form
$R={\cal U}_{1} D {\cal U}^{\small T}_{2},$
where
${\cal U}_{1}$ and ${\cal U}_{2}$  are
real orthogonal $(N-1)\times(N-1)$ matrices and
$D=diag(\lambda_1,...,\lambda_{N-1})$
is diagonal $(N-1)\times(N-1)$  matrix
such that
\quad $\lambda_1 \ge \lambda_2 \ge ... \ge \lambda_{N-1} \ge 0$.
}

\noindent {\bf Proof.}
This theorem is proved  in \cite{EY,Lan,Sc,Gant}.

\noindent {\bf Theorem 2.} (Singular Valued Decomposition for Gates)\\
{\it Any linear quantum four-valued logic gate (\ref{LGE})
can be represented by
\[ \hat{\cal E}=\hat{\cal E}^{(T)} \hat{\cal U}_{1} \
\hat D \ \hat{\cal U}_{2}, \]
where\\
$\hat{\cal U}_{1}$ and $\hat{\cal U}_{2}$ are
unital orthogonal quantum gates
\[ \hat{\cal U}_{i}=|0)(0|+\sum^{N-1}_{\mu=1} \sum^{N-1}_{\nu=1}
R^{(i)}_{\mu \nu} |\mu)(\nu| , \]
$\hat D$ is a unital diagonal quantum gate, such that
\[ \hat D=|0)(0|+\sum^{N-1}_{\mu=1}
\lambda_{\mu} |\mu)(\mu|, \]
where $\lambda_{\mu} \ge 0$.\\
$\hat{\cal E}^{(T)}$ is a translation quantum gate, such that
\[ \hat{\cal E}^{(T)}=|0)(0|+\sum^{N-1}_{\mu=1}
|\mu)(\mu|+\sum^{N-1}_{\mu=1} T_{\mu} |\mu)(0|. \]
}

\noindent {\bf Proof.} The proof of this theorem can be easy realized
in matrix representation by using Lemma 3 and Theorem 1.

As a result we have that any trace-preserving gate can be realized
by 3 types of gates: (1) unital orthogonal quantum gates
$\hat{\cal U}$ with matrix ${\cal U}\in SO(4^{n}-1,\mathbb{R})$; (2)
unital diagonal quantum gate $\hat D$ with matrix $D \in
D(4^{n}-1,\mathbb{R})$; (3) nonunital translation gate $\hat{\cal
E}^{(T)}$ with matrix ${\cal E}^{(T)}\in T(4^{n}-1,\mathbb{R})$.

%%%%%%%%%%%%%%%%%%%%%%%%%%%%%%%%%%%%%%%%%%%%%%%%%%%%%%%%%%%%%%%%%%%%%%%%%

\section{Universal set of quantum gates }

%%%%%%%%%%%%%%%%%%%%%%%%%%%%%%%%%%%%%%%%%%%%%%%%%%%%%%%%%%%%%%%%%%%%%%%%%

The condition for performing arbitrary unitary operations on pure state to
realize a quantum computation by unitary dynamics 
is well understood  \cite{Bar,DBE,DV,LL}. Using a universal
gate set, quantum computations may realize the time sequence of
operations corresponding to any unitary dynamics. Deutsch, Barenco
and Ekert \cite{DBE}, DiVincenzo \cite{DV} and Lloyd \cite{LL}
showed that almost any two-qubits quantum gate is universal. It is
known \cite{Bar,DBE,DV,LL} that a set of quantum gates that
consists of all one-qubit gates and the two-qubits exclusive-or 
gate is universal in the sense that all unitary operations
on arbitrary many qubits can be expressed as compositions of these
gates. Recently in the paper \cite{BB} was considered universality
for n-qudits quantum gates.

The same is not true for the general quantum operations. 
In the paper \cite{Bac} single qubit
quantum system with Markovian dynamics was considered and
the resources needed for universality of general quantum
operations was studied.
An analysis of completely-positive trace-preserving superoperators on
single qubit density matrices was realized in papers \cite{FA,KR,RSW}.

Let us study universality for general quantum four-valued logic gates.
A set of quantum four-valued logic gates is universal iff all
quantum gates on arbitrary many ququats can be expressed
as compositions of these gates.
A set of quantum four-valued logic gates is universal iff all unitary
two-valued logic gates and general quantum operations
can be represented by compositions of these gates.
Single ququat gates cannot map two initially
un-entangled ququats into an entangled state.
Therefore the single ququat gates or set of single ququats gates
are not universal gates.
Quantum gates which are realization of classical gates
cannot be universal by definition, since these gates evolve
generalized computational states to generalized computational states
and never to the superposition of them.

Let us consider linear completely positive
trace-decreasing superoperator $\hat{\cal E}$.
This superoperator can be represented in the form (\ref{elr}),
where $\hat L_{A}$ and $\hat R_{A}$ are 
superoperators on $\overline{\cal H}^{(n)}$ defined by
$\hat L_{A}|B)=|AB)$ and $\hat R_{A}|B)=|BA)$.

The n-ququats linear gate $\hat{\cal E}$ is completely positive
trace-preserving superoperator such that the gate matrix is an
element of Lie group $TGL(4^{n}-1,\mathbb{R})$. In general case, the
n-ququats nonlinear gate $\hat{\cal N}$ is defined by completely
positive trace-decreasing linear superoperator $\hat{\cal E}$ such
that the gate matrix is an element of Lie group $GL(4^{n},{\bf
R})$. The condition of completely positivity leads to difficult
inequalities for gate matrix elements \cite{Choi,FA,KR,RSW}. In
order to satisfy condition of completely positivity we use the
representation (\ref{elr}). {\it To find the universal set of
completely positive (linear or nonlinear) gates $\hat{\cal E}$ we
consider the universal set of the superoperators $\hat L_{A_j}$
and $\hat R_{A^{\dagger}_j}$.} The matrices of superoperators 
$\hat L_A$ and $\hat R_A$ are
connected by complex conjugation. Obviously, the universal set of
these superoperators defines 
a universal set of completely
positive superoperators $\hat{\cal E}$ of the quantum gates.

Let the superoperators $\hat L_{A}$ and $\hat R_{A^{\dagger}}$
be called pseudo-gates. These superoperators can be represented by
\[ \hat L_A=  \sum^{N-1}_{\mu=0}\sum^{N-1}_{\nu=0} 
L^{(A)}_{\mu \nu} |\mu)(\nu|, \quad
\hat R_{A^{\dagger}}= \sum^{N-1}_{\mu=0}  \sum^{N-1}_{\nu=0}
R^{(A^{\dagger})}_{\mu \nu} |\mu)(\nu|. \]

%%{\bf Lemma 4.}\\
\bp
{\it The matrix ${\cal E}_{\mu \nu}$ of the completely 
positive superoperator (\ref{elr})
can be represented by}
\begin{equation} \label{ELR-M} {\cal E}_{\mu \nu}
=\sum^{m}_{j=1} \sum^{N-1}_{\alpha=0} L^{(jA)}_{\mu \alpha}
R^{(jA^{\dagger})}_{\alpha \nu} \ . \end{equation}
\ep

\noindent {\bf Proof.}
This Lemma can be easy  prooved in matrix representation \cite{Tarpr}.

The matrix elements $L^{(jA)}_{\mu \alpha}$ 
and $R^{(jA^{\dagger})}_{\alpha \nu}$ can be rewritten in the form
\begin{equation} \label{LRmat} L^{(jA)}_{\mu \alpha}=\frac{1}{2^n}
(\sigma_{\mu} \sigma_{\alpha}| A) \ , \quad
R^{(jA^{\dagger})}_{\alpha \nu}=
%%%\frac{1}{2^n}(\sigma_{\nu} \sigma_{\alpha}| A^{\dagger})=
\frac{1}{2^n}(A|\sigma_{\alpha} \sigma_\nu). \end{equation}

Using 
\[ (L^{(jA)}_{\mu \alpha})^{*}= \frac{1}{2^n}
(\sigma_{\mu} \sigma_{\alpha}| A)^{*}=
\frac{1}{2^n} (A|\sigma_{\mu} \sigma_{\alpha})=
R^{(jA^{\dagger})}_{\mu \alpha}. \]
we get the matrices
$L^{(jA)}_{\mu \alpha}$ and $R^{(jA^{\dagger})}_{\mu \alpha}$
are complex $4^{n}\times 4^{n}$ matrices and their elements
are connected by complex conjugation: 
$(L^{(jA)}_{\mu \alpha})^{*}=R^{(jA^{\dagger})}_{\mu \alpha}$.
We can write the gate matrix (\ref{ELR-M}) in the form
\[ {\cal E}_{\mu \nu}
=\sum^{m}_{j=1} \sum^{N-1}_{\alpha=0} L^{(jA)}_{\mu \alpha}
(L^{(jA)}_{\alpha \nu})^{*}. \]

A two-ququats gate $\hat{\cal E}$ is called primitive \cite{BB} if
$\hat{\cal E}$ maps tensor product of single ququats
to tensor product of single ququats, i.e.
if $|\rho_{1})$ and $|\rho_{2})$ are ququats, then
we can find ququats $|\rho^{\prime}_{1})$ and $|\rho^{\prime}_{2})$
such that
$\hat{\cal E}|\rho_{1}) \otimes\rho_{2})=
|\rho^{\prime}_{1} \otimes \rho^{\prime}_{2})$.
The superoperator $ \hat{\cal E}$ is called imprimitive if
$\hat{\cal E}$ is not primitive.

It can be shown that almost every pseudo-gate that operates
on two or more ququats is universal pseudo-gate.

{\bf Theorem 3.}\\
{\it The set of all single ququat pseudo-gates and
any imprimitive two-ququats pseudo-gate
are universal set of pseudo-gates.}

\noindent {\bf Proof.} 
Expressed in  group theory language, all n-ququats pseudo-gates
are elements of the Lie group $GL(4^n,\mathbb{C})$. Two-ququats
pseudo-gates $\hat L$ are elements of Lie group $GL(16,\mathbb{C})$.
The question of universality is the same as the question of what
set of superoperators $\hat L$ sufficient to generate $GL(16,\mathbb{C})$.
The group $GL(16,\mathbb{C})$ has $(16)^2=256$
independent one-parameter subgroups $GL_{\mu \nu}(16,\mathbb{C})$
of one-parameter pseudo-gates $\hat L^{(\mu \nu)}(t)$ such that
$\hat L^{(\mu \nu)}(t)=t|\mu)(\nu|$. 
Infinitesimal generators of Lie group $GL(4^n,\mathbb{C})$ are defined by
\[ \hat H_{\mu \nu}=\Bigl(\frac{d}{dt} \hat L^{(\mu \nu)}(t) \Bigr)_{t=0}, \]
where $\mu,\nu=0,1,...,4^{n}-1$.
The generators $\hat H_{\mu \nu}$ of the one-parameter subgroup
$GL_{\mu \nu}(4^n,\mathbb{R})$ are superoperators of the form
$\hat H_{\mu \nu}=|\mu)(\nu|$ on $\overline{\cal H}^{(n)}$ 
which can be represented by
$4^n \times 4^n$ matrix $H_{\mu \nu}$ with elements
$(H_{\mu \nu})_{\alpha \beta}=\delta_{\alpha \mu} \delta_{\beta \nu}$.
The set of superoperators $\hat H_{\mu \nu}$ is a basis 
(Weyl basis \cite{BR})
of Lie algebra $gl(16,\mathbb{R})$ such that
\[ [\hat H_{\mu \nu},\hat H_{\alpha \beta}]=
\delta_{\nu \alpha} \hat H_{\mu \beta}- \delta_{\mu \beta} \hat
H_{\nu \alpha}, \] 
where $\mu, \nu, \alpha, \beta =0,1,...,15.$
Any element $\hat H$ of the algebra
$gl(16,\mathbb{C})$ can be represented by
\[ \hat H=\sum^{15}_{\mu=0}\sum^{15}_{\nu=0}
h_{\mu \nu} \hat H_{\mu \nu}, \]
where $h_{\mu \nu}$ are complex coefficients.

As a basis of Lie algebra $gl(16,\mathbb{C})$ 
we can use $256$ linearly independent
self-adjoint superoperators
\[ H_{\alpha \alpha}=|\alpha)(\alpha|, \quad
H^r_{\alpha \beta}=|\alpha)(\beta|+|\beta)(\alpha|,\]
\[ H^i_{\alpha \beta}= -i\Bigl(
|\alpha)(\beta|-|\beta)(\alpha|\Bigr). \]
where $0\le \alpha \le \beta \le 15$.
The matrices of these generators is
Hermitian $16 \times 16$ matrices.
The matrix elements of 256 Hermitian $16 \times 16$ matrices
$H_{\alpha \alpha}$,  $H^r_{\alpha \beta}$ and $H^{i}_{\alpha \beta}$
are defined by
\[ (H_{\alpha \alpha})_{\mu \nu}=
\delta_{\mu \alpha} \delta_{\nu \alpha} \ ,
\quad (H^r_{\alpha \beta})_{\mu \nu}=
\delta_{\mu \alpha} \delta_{\nu \beta}
+\delta_{\mu \beta} \delta_{\nu \alpha}, \]
\[ (H^i_{\alpha \beta})_{\mu \nu}=
-i(\delta_{\mu \alpha} \delta_{\nu \beta}
-\delta_{\mu \beta} \delta_{\nu \alpha}). \]
For any Hermitian generators $\hat H$ exists
one-parameter pseudo-gates $\hat L(t)$ 
which can be represented in the form 
$\hat L(t)=exp \ it \hat H$ such that 
$\hat L^{\dagger}(t)\hat L(t)=\hat I$.
%%Linear superoperators $\hat H$ are
%%infinitesimal generators of Lie group.

Let us write main operations which allow to derive
new pseudo-gates $\hat L$ from a set of pseudo-gates.\\
1) We introduce general SWAP (twist)
pseudo-gate $\hat T^{(SW)}$.
A new pseudo-gate $\hat L^{(SW)}$ defined by
$\hat L^{(SW)}=\hat T^{(SW)} \hat L \hat T^{(SW)}$
is obtained directly from $\hat L$ by exchanging two ququats.\\
2) Any superoperator $\hat L$ on $\overline{\cal H}^{(2)}$
generated by the commutator
$i[\hat H_{\mu\nu}, \hat H_{\alpha \beta}]$ can be obtained
from $\hat L_{\mu\nu}(t)=exp \ it\hat H_{\mu\nu}$
and $\hat L_{\alpha \beta}(t)=exp \ it\hat H_{\alpha \beta}$ because
\[ exp \ t \ [\hat H_{\mu\nu},\hat H_{\alpha \beta}]=\]
\[=\lim_{n \rightarrow \infty} \Bigl( \hat L_{\alpha \beta}(-t_n)
\hat L_{\mu\nu}(t_n) \hat L_{\alpha \beta}(t_n) \hat
L_{\mu\nu}(-t_n)\Bigr)^n, \]
where $t_n=1/\sqrt{n}$.
Thus we can use the commutator
$i[\hat H_{\mu \nu}, \hat H_{\alpha \beta}]$
to generate pseudo-gates.\\
3) Every transformation $\hat L(a,b)=exp i\hat H(a,b)$
of $GL(16,\mathbb{C})$ generated by
superoperator $\hat H(a,b)=a\hat H_{\mu\nu}+b\hat H_{\alpha \beta}$, where
a and b is complex, can obtained from
 $\hat L_{\mu\nu}(t)=exp \ it\hat H_{\mu\nu}$
and $\hat L_{\alpha \beta}(t)=exp \ it\hat H_{\alpha \beta}$   by
 \[ exp \ i \hat H(a,b)=
\lim_{n \rightarrow \infty} \Bigl( \hat L_{\mu \nu}(\frac{a}{n})
\hat L_{\alpha \beta}(\frac{b}{n})\Bigr)^n. \]

For other details of the proof, see \cite{DV,DBE,BB,Bar,LL}.

%%%%%%%%%%%%%%%%%%%%%%%%%%%%%%%%%%%%%%%%%%%%%%%%%%%%%%%%%%%%%%%%%%%%%%%%
\section{Examples of general quantum gates}
%%%%%%%%%%%%%%%%%%%%%%%%%%%%%%%%%%%%%%%%%%%%%%%%%%%%%%%%%%%%%%%%%%%%%%%%%

\subsection{Unitary quantum gates}

Let us use Lemma 1.
In the generalized computational basis any unitary two-valued logic
gate $U$ can be considered as a quantum four-valued logic gate:
\begin{equation} \label{P6-1}
\hat{\cal U}=\sum^{N-1}_{\mu=0}\sum^{N-1}_{\nu=0}
{\cal U}_{\mu \nu} |\mu)(\nu|  \ , \end{equation}
where ${\cal U}_{\mu \nu}$ is a real matrix such that
\begin{equation} \label{P6-2}
{\cal U}_{\mu \nu}=\frac{1}{2^n}Tr\Bigl(\sigma_{\nu}U
\sigma_{\mu} U^{\dagger}\Bigr) \ . \end{equation}
This formula defines a relation between unitary quantum two-valued
logic gates $U$ and the real $4^{n}\times 4^{n}$ matrix ${\cal U}$.

Any four-valued logic gate associated with unitary 2-valued
logic gate by (\ref{P6-1},\ref{P6-2}) is unital gate, i.e.
gate matrix ${\cal U}$ defined by (\ref{P6-2}) has
${\cal U}_{\mu 0}={\cal U}_{0 \mu }=\delta_{\mu 0}$.
\[ {\cal U}_{\mu 0}=\frac{1}{2^{n}}
Tr\Bigl(\sigma_{\mu}U\sigma_{0} U^{\dagger}\Bigr)=\frac{1}{2^{n}}
Tr\Bigl(\sigma_{\mu}UU^{\dagger}\Bigr)=\frac{1}{2^{n}}
Tr\sigma_{\nu}. \]
Using $Tr\sigma_{\mu}=\delta_{\mu 0}$
we get ${\cal U}_{\mu 0}=\delta_{\mu 0}$.

Let us denote the gate $\hat{\cal U}$ associated with unitary
two-valued logic gate $U$ by $\hat{\cal E}^{(U)}$.
%%{\bf Lemma 5.}\\
\bp
{\it If $U$ is unitary two-valued logic gate, then
in the generalized computational basis a quantum four-valued
logic gate $\hat{\cal U}=\hat{\cal E}^{(U)}$ associated with $U$
is represented
by orthogonal matrix ${\cal E}^{(U)}$: }
\begin{equation} \label{ORT}
{\cal E}^{(U)}({\cal E}^{(U)})^{T}=
({\cal E}^{(U)})^{T}{\cal E}^{(U)}=I \ .
\end{equation}
\ep

\noindent {\bf Proof.}
Let $\hat{\cal E}^{(U)}$ is defined by
\[ \hat{\cal E}^{(U)}\rho=U \rho U^{\dagger} \ ,
\quad \hat{\cal E}^{(U^{\dagger})}\rho=U^{\dagger} \rho U. \]
If $UU^{\dagger}=U^{\dagger}U=I$, then
$\hat{\cal E}^{(U)}\hat{\cal E}^{(U^{\dagger})}=
\hat{\cal E}^{(U^{\dagger})}\hat{\cal E}^{(U)}=\hat I$. 
In the matrix representation we have
\[ \sum^{N-1}_{\alpha=0} {\cal E}^{(U)}_{\mu \alpha}
{\cal E}^{(U^{\dagger})}_{\alpha \nu}=
\sum^{N-1}_{\alpha=0} {\cal E}^{(U^{\dagger})}_{\mu \alpha}
{\cal E}^{(U)}_{\alpha \nu}=\delta_{\mu \nu}  \ , \]
i.e. ${\cal E}^{(U^{\dagger})}{\cal E}^{(U)}=
{\cal E}^{(U)}{\cal E}^{(U^{\dagger})}=I$.
Note that
\[ {\cal E}^{(U^{\dagger})}_{\mu \nu}=
\frac{1}{2^{n}}Tr\Bigl( \sigma_{\mu} U^{\dagger}\sigma_{\nu} U \Bigr)=
\frac{1}{2^{n}}Tr\Bigl( \sigma_{\nu} U \sigma_{\mu} U^{\dagger} \Bigr)=
{\cal E}^{(U)}_{\nu \mu}, \]
%%%=({\cal E}^{(U)}_{\mu \nu})^{T}, \]
i.e. ${\cal E}^{(U^{\dagger})}=({\cal E}^{(U)})^{T}$.
Finally, we obtain (\ref{ORT}).

Note that n-qubit unitary two-valued logic gate
$U$ is an element of Lie group $SU(2^{n})$.
The dimension of this group is equal to
$dim \ SU(2^{n})=(2^{n})^{2}-1=4^{n}-1$.
The matrix of n-ququat orthogonal linear gate
$\hat{\cal U}=\hat{\cal E}^{(U)}$ can be considered as an element
of Lie group $SO(4^{n}-1)$.
The dimension of this group is equal to
$dim \ SO(4^{n}-1)=(4^{n}-1)(2 \cdot 4^{n-1}-1)$.
For example, if $n=1$, then
$dim \ SU(2^{1})=dim \ SO(4^{1}-1)=3$.
If $n=2$, then
$dim \ SU(2^{2})=15$, $dim \  SO(4^{2}-1)=105$. 
Therefore not all orthogonal 4-valued logic gates for mixed
and pure states are connected
with unitary 2-valued logic gates for pure states.

Let us consider single ququat 4-valued logic gate $\hat{\cal U}$
associated with unitary single qubit 2-valued logic gate $U$.

%%{\bf Lemma 6.}\\
\bp {\it Any single-qubit unitary quantum two-valued logic gate
can be realized as the product of single ququat simple rotation
gates $\hat{\cal U}^{(1)}(\alpha)$, $\hat{\cal U}^{(2)}(\theta)$
and $\hat{\cal U}^{(1)}(\beta)$ defined by
\[ \hat{\cal U}^{(1)}(\alpha)=|0)(0|+|3)(3|+
\cos \alpha \Bigl(|1)(1|+|2)(2| \Bigr)+\]
\[+\sin \alpha \Bigl(|2)(1|-|1)(2| \Bigr), \]
\[ \hat{\cal U}^{(2)}(\theta)=|0)(0|+|2)(2|+
\cos \theta \Bigl(|1)(1|+|3)(3| \Bigr)+\]
\[+\sin \theta \Bigl(|1)(3|-|3)(1| \Bigr), \]
where $\alpha$, $\theta$ and $\beta$ are Euler angles.
}
\ep

\noindent {\bf Proof.} See \cite{Tarpr}. 

\noindent{\bf Example 1.}
In the generalized computational basis the unitary NOT gate
("negation") of two-valued logic
\[ X=|0><1|+|1><0|=\sigma_{1}, \]
is represented by quantum four-valued logic gate
\[ \hat{\cal E}^{(X)}=|0)(0|+|1)(1|-|2)(2|-|3)(3|. \]

\noindent{\bf Example 2.}
The Hadamar two-valued logic gate
\[ H=\frac{1}{\sqrt{2}}(\sigma_1+\sigma_3)  \]
can be represented as a four-valued logic gate by
\[ \hat{\cal E}^{(H)}=|0)(0|-|2)(2|+|3)(1|+|1)(3|. \]

\subsection{Measurements as quantum gates}

It is known that von Neumann measurement operation ${\cal E}$ is
\begin{equation} \label{vNm}
{\cal E}(\rho)=\sum^{r}_{k=1} P_{k}\rho P_{k} \ , \end{equation}
where $\{P_{k}|k=1,..,r\}$ is a (not necessarily complete)
sequence of orthogonal projection operators on ${\cal H}^{(n)}$.

Let $P_{k}$ are projectors onto the pure state $|k>$
which define usual computational basis $\{|k>\}$, i.e.
$P_{k}=|k><k|$. 

%%{\bf Lemma 7.}\\
\bp
{\it A nonlinear four-valued logic gate $\hat{\cal N}$
for von Neumann measurement (\ref{vNm}) 
of the state $\rho=\sum^{N-1}_{\alpha=0}|\alpha) \rho_{\alpha}$
is defined by
\[ \hat{\cal N}=\sum^{r}_{k=1} \frac{1}{p(k)}
{\cal E}^{(k)}_{\mu \nu} |\mu)(\nu|, \]
where
\begin{equation} \label{EE1}
{\cal E}^{(k)}_{\mu \nu}=
%%(\mu|\hat{\cal E}_{k}|\nu)=
\frac{1}{2^{n}} Tr(\sigma_{\mu} P_{k} \sigma_{\nu} P_{k}),
\quad p(k)=\sqrt{2^n} \sum^{N-1}_{\alpha=0} {\cal E}^{(k)}_{0 \alpha}
\rho_{\alpha} \ . \end{equation}
}
\ep

\noindent {\bf Proof.}
The trace-decreasing quantum operation ${\cal E}_{k}$ is defined by
$\rho \ \rightarrow \ \rho^{\prime}={\cal E}_{k}(\rho)=P_{k}\rho P_{k}$. 
The superoperator $\hat{\cal E}$ for this quantum operation has
the form $\hat{\cal E}_{k}=\hat L_{P_{k}} \hat R_{P_{k}}$ and
$|\rho')=\hat{\cal E}_{k}|\rho)$. Then
\[ {\rho'}_{\mu}=(\mu|\rho')=(\mu|\hat{\cal E}_{k}|\rho)=
\sum^{N-1}_{\nu=0} (\mu|\hat{\cal E}_{k}|\nu)(\nu|\rho)=
\sum^{N-1}_{\nu=0} {\cal E}^{(k)}_{\mu \nu} \rho_{\nu}\ , \]
where
${\cal E}^{(k)}_{\mu \nu}=(\mu|\hat{\cal E}_{k}|\nu)=
(1/2^{n}) Tr(\sigma_{\mu} P_{k} \sigma_{\nu} P_{k})$. 
The probability that process represented by $\hat{\cal E}_{k}$ 
occurs is
\[ p(k)=Tr(\hat{\cal E}_k(\rho))=(I|\hat{\cal E}|\rho)=
\sqrt{2^n}\rho^{\prime}_0=\sqrt{2^n}
\sum^{N-1}_{\alpha=0} {\cal E}^{(k)}_{0 \alpha} \rho_{\alpha} \ . \]
If
$\sum^{N-1}_{\alpha=0} {\cal E}_{0 \alpha} \rho_{\alpha} \not=0 $,
then the matrix for nonlinear trace-preserving gate $\hat{\cal N}$ is
\[ {\cal N}_{\mu \nu}=\sqrt{2^n}
(\sum^{N-1}_{\alpha=0}{\cal E}_{0 \alpha} \rho_{\alpha})^{-1}
{\cal E}_{\mu \nu}. \]

\noindent{\bf Example.}
Let us consider single ququat projection operator
$P_{+}=|0><0|$ and $P_{-}=|1><1|$ which can be defined by
\[ P_{\pm}=\frac{1}{2}(\sigma_{0} \pm \sigma_{3}). \]
Using formula (\ref{EE1}) we derive
\[ {\cal E}^{(\pm)}_{\mu \nu}=
\frac{1}{8}Tr\Bigl(\sigma_{\mu}(\sigma_{0} \pm \sigma_{3})
\sigma_{\nu}(\sigma_{0} \pm \sigma_{3}) \Bigr)=\]
\[=\frac{1}{2}\Bigl(
\delta_{\mu 0}\delta_{\nu 0}+\delta_{\mu 3} \delta_{\nu 3} \pm
\delta_{\mu 3}\delta_{\nu 0} \pm \delta_{\mu 0} \delta_{\nu 3}
\Bigr). \]
The linear trace-decreasing superoperator
for von Neumann measurement
projector $P_{\pm}$ onto pure state is
\[ \hat{\cal E}^{(\pm)}=\frac{1}{2}\Bigl( |0)(0)+|3)(3| \pm
|0)(3| \pm |3)(0| \Bigr). \]
The superoperators $\hat{\cal E}^{(\pm)}$ are not trace-preserving.
The probabilities $p_{\pm}$ that processes represented by superoperators
$\hat{\cal E}^{(\pm)}$ occurs are
\[ p_{\pm}=\frac{1}{\sqrt{2}}(\rho_0+\rho_{3}). \]

%%%%%%%%%%%%%%%%%%%%%%%%%%%%%%%%%%%%%%%%%%%%%%%%%%%%%%%%%%%%%%%%%%%
\subsection{Quantum gates for classical gates}

For the concept of many-valued logic see Appendix II and \cite{Re,RT,Ya2}.

Let us consider linear trace-preserving quantum gates for
classical gates $\sim$, $\overline{x}$,
$I_{0},I_{1},I_{2},I_{3}$, $0,1,2,3$,
 $\diamondsuit$, $\Box$.

%%{\bf Lemma 8.}\\
\bp
{\it Any single argument classical gate $g(\nu)$
can be realized as linear trace-preserving quantum
four-valued logic gate by
\[ \hat{\cal E}^{(g)}=|0)(0|+\sum^{3}_{k=1 }|g(k))(k|+\]
\[+(1-\delta_{0g(0) })\Bigl( |g(0))(0|-
\sum^{3}_{\mu=0} \sum^{3}_{\nu=0}(1-\delta_{\mu g(\nu)})
|\mu)(\nu| \Bigr). \]
}
\ep

\noindent {\bf Proof.}
The proof is by direct calculation in
$\hat{\cal E}^{(g)}|\alpha]=|g(\alpha)]$, 
where
$\hat{\cal E}^{(g)}|\alpha]= (1/\sqrt{2})\Bigl(
\hat{\cal E}^{(g)}|0)+\hat{\cal E}^{(g)}|\alpha) \Bigr)$. 

\noindent{\bf Examples.}

1. Luckasiewicz negation gate is
\[ \hat{\cal E}^{(\sim x)}=|0)(0|+|1)(2|+|2)(1|+|3)(0|-|3)(3|. \]

2.  The four-valued logic gate $I_{0}$ can be realized by
\[ \hat{\cal E}^{(I_0)}=|0)(0|+|3)(0|-\sum^{3}_{k=1}|3)(k|. \]

3. The gates $I_{k}(x)$, where $k=1,2,3$ are
\[ \hat{\cal E}^{(I_k)}=|0)(0|+|3)(k|. \]

4.  The gate $\overline{x}$ can be realized by
\[ \hat{\cal E}^{(\overline{x})}=|0)(0|+|1)(0|+|2)(1|+|3)(2|-
\sum^{3}_{k=1}|1)(k|. \]

5. The constant gates $0$ and $k=1,2,3$ can be realized by
\[ \hat{\cal E}^{(0)}=|0)(0|  \ , \quad
\hat{\cal E}^{(k)}=|0)(0|+|k)(0|. \]

6.  The gate $\diamondsuit x$ is realized by
\[ \hat{\cal E}^{(\diamondsuit)}=|0)(0|+\sum^{3}_{k=1}|3)(k|. \]

7. The gate $\Box x= \sim \diamondsuit x$ is
$\hat{\cal E}^{(\Box)}=|0)(0|+|3)(3|$. 

\noindent
Note that quantum gates $\hat{\cal E}^{(\sim x)}$,
$\hat{\cal E}^{(I_{0})}$, $\hat{\cal E}^{(k)}$ 
are not unital gates.

%%\subsection{Quantum gates for two-arguments classical gates}

Let us consider quantum gates for two-arguments classical gates.

1. The generalized conjunction $x_{1} \land x_{2}=min(x_{1},x_{2})$ and
generalized disjunction $x_{1} \land x_{2}=max(x_{1},x_{2})$ can be
realized by two-ququats unital gate:
\[\hat{\cal E}|x_1,x_2]=|x_{1} \lor x_{2},x_{1} \land x_{2}] \ . \]

Let us write the quantum gate
which realizes the these classsical gates in the generalized computational basis by
\[ \hat{\cal E}=\sum^{N-1}_{\mu}  \sum^{N-1}_{\nu} |\mu\nu)(\mu \nu|+
\sum^{3}_{k=1}\Bigl( |0k)-|k0) \Bigr)(k 0|+\]
\[ +\sum^{3}_{k=2}\Bigl( |1k)-|k1) \Bigr)(k 1|+
\Bigl( |23)-|32) \Bigl)(32|. \]

2.  The Sheffer-Webb function gate 
\[ \hat{\cal E}|x_{1},x_{2}]=
|V_{4}(x_{1},x_{2}),\sim V_{4}(x_{1},x_{2})] \] 
can be realized by  two-ququats gate:
\[ \hat{\cal E}=|00)(00|+|12)(00|-
\sum^{3}_{\mu=0}\sum^{3}_{\nu=1}|12)(\mu \nu|+|21)(10|+ \]
\[ +|21)(11|+|30)(02|+|30)(20|+|30)(12|+|30)(21|+\]
\[ +|30)(22|+|03)(03|+|03)(13|+|03)(23|+\sum^{3}_{\mu=0} |03)(3\mu|. \]
Note that this gate is not unital quantum gate.

%%%%%%%%%%%%%%%%%%%%%%%%%%%%%%%%%%%%%%%%%%%%%%%%%%%%%%%%%%%%%%%%%%%%%%%%%

\section{Appendix I.}

%%%%%%%%%%%%%%%%%%%%%%%%%%%%%%%%%%%%%%%%%%%%%%%%%%%%%%%%%%%%%%%%%%%%%%%%%%

\subsection{Pure states and Hilbert space}
%%{\bf 2.1. Pure states}\\

A quantum system in a pure state is described by unit vector in a
Hilbert space ${\cal H}$. In the Dirac notation a pure state is
denoted by $|\Psi>$. The Hilbert space ${\cal H}$ is a linear
space with an inner product. The inner product  for $ |\Psi_{1}
>$, $|\Psi_{2}> \in {\cal H}$ is denoted by $<\Psi_{1}|\Psi_{2}>
$. A quantum bit or qubit, the fundamental concept of quantum
computations, is a two-state quantum system. The two basis states
labeled $|0>$ and $|1>$, is orthogonal unit vectors, i.e.
$<k|l>=\delta_{kl}$, where $k,l\in \{0,1\}$.
The Hilbert space of qubit is ${\cal H}_{2}={\mathbb{C}}^2$.
The quantum system which is used to quantum computations 
consists of n quantum two-state particles.
The Hilbert space ${\cal H}^{(n)}$ of such a system is a tensor product
of n Hilbert spaces ${\cal H}_{2}$ of one two-state particle: \
${\cal H}^{(n)}={\cal H}_{2}
\otimes {\cal H}_{2} \otimes ... \otimes {\cal H}_{2}$.
The space ${\cal H}^{(n)}$ is a $N=2^{n}$ dimensional complex linear
space. Let us choose a basis for ${\cal H}^{(n)}$
which is consists of the $N=2^{n}$ orthonormal states
$|k>$, where k is in binary representation.
The state is a tensor product of states in ${\cal H}^{(n)}$:
\[ |k>=|k_{1}>\otimes |k_{2}> \otimes ... \otimes |k_{n}>=
|k_{1}k_{2}...k_{n}>  \ , \]
where $k_{i} \in \{0,1\}$ and $i=1,2,...,n$.
This basis is usually called computational basis
which has $2^{n}$ elements.
A pure state $|\Psi(t)> \in {\cal H}^{(n)}$ is generally a superposition
of the basis states
\begin{equation} \label{Psi} |\Psi(t)>=\sum^{N-1}_{k=0}a_{k}(t)|k> \ ,
\end{equation}
with $N=2^{n}$ and $\sum^{N-1}_{k=0} |a_{k}(t)|^{2}=1$.

%%%%%%%%%%%%%%%%%%%%%%%%%%%%%%%%%%%%%
\subsection{Operator Hilbert space}

For the concept of operator Hilbert space and superoperators see
\cite{Cra}-\cite{kn2}.

The space of linear operators acting on a
$N=2^{n}$-dimensional Hilbert space ${\cal H}^{(n)}$
is a $N^{2}=4^{n}$-dimensional complex linear space
$\overline {\cal H}^{(n)}$.
We denote an element $A$ of $\overline{\cal H}^{(n)}$
by a ket-vector $|A)$. The inner product of two elements $|A)$ and
$|B)$ of $\overline{\cal H}^{(n)}$ is defined as
\begin{equation} \label{inner} (A|B)=Tr(A^{\dagger} B) \ . \end{equation}
The norm $\|A\|=\sqrt{(A|A)}$ is the Hilbert-Schmidt norm of
operator $A$. A new Hilbert space $\overline{\cal H}$ with scalar
product (\ref{inner}) is called operator Hilbert space attached to ${\cal
H}$ or the associated Hilbert space, or Hilbert-Schmidt space
\cite{Cra}-\cite{kn2}.

Let $\{|k>\}$ be an orthonormal basis of ${\cal H}^{(n)}$:
\[ <k|k'>=\delta_{kk'} \ , \quad \sum^{N-1}_{k=0}|k><k|=I. \]
Then $|k,l)=||k><l|)$
is an orthonormal basis of the operator Hilbert space $\overline{\cal H}^{(n)}$:
\[ (k,l|k',l')=\delta_{kk'}\delta_{ll'} \ , \quad
\sum^{N-1}_{k=0} \sum^{N-1}_{l=0}|k,l)(k,l|=\hat I , \]
where $N=2^n$. The operator basis $|k,l)$ has $4^{n}$ elements.
Note that $|k,l) \not= |kl>=|k>\otimes |l>$ and
\[ |k,l)=|k_{1},l_{1})\otimes |k_{2},l_{2}) \otimes ...
\otimes |k_{n},l_{n}) , \]
where $k_{i},l_{i} \in \{0,1\}$, $i=1,...,n$ and
\[ |k_{i},l_{i})\otimes |k_{j},l_{j})=
| \ |k_{i}>\otimes  |k_{j}>, <l_{i}| \otimes <l_{j}| \ ). \]
For an arbitrary element $|A)$ of $\overline{\cal H}^{(n)}$ we have
\[ |A)=\sum^{N-1}_{k=0}\sum^{N-1}_{l=0} |k,l)(k,l|A) \]
with $(k,l|A)=Tr( |l><k| A )=<k|A|l>=A_{kl}$.
An operator $\rho$ of density matrix for n-qubits can be considered
as an element $|\rho)$ of space $\overline{\cal H}^{(n)}$.

\subsection{Superoperators}

Operators, which act on $\overline{\cal H}$, are called superoperators
and we denote them in general by the hat.

For an arbitrary superoperator $\hat \Lambda$ on
$\overline{\cal H}$, which is defined by
\[ \hat \Lambda|A)=|\Lambda(A) ), \]
we have
\[ (k,l|\hat \Lambda|A)=\sum^{N-1}_{k'=0}\sum^{N-1}_{l'=0}
(k,l|\hat \Lambda|k',l') (k',l'|A)=\]
\[ =\sum^{N-1}_{k'=0}\sum^{N-1}_{l'=0}
\Lambda_{klk'l'}A_{k'l'}, \]
where $N=2^n$.

Let $A$ be a linear operator in Hilbert space.
Then the superoperators $\hat L_{A}$ and $\hat R_{A}$ will be defined by
\[ \hat L_{A}|B)=|AB) \ , \quad \hat R_{A}|B)=|BA). \]
The superoperator $\hat P=|A)(B|$ is defined by
\[ \hat P|C)=|A)(B|C)=|A)Tr(B^{\dagger}C). \]

A superoperator $\hat{\cal E}^{\dagger}$ is called the adjoint
superoperator for $\hat{\cal E}$ if
\[ (\hat{\cal E}^{\dagger}(A)|B)=(A|\hat{\cal E}(B)) \]
for all $|A)$ and $|B)$ from $\overline{\cal H}$.
A superoperator $\hat{\cal E}$ is unital if $\hat{\cal E}(I)=I$.

Pauli matrices can be considered as a basis in
operator space. Let us write the Pauli matrices in
the form
%$$
%\sigma_{1}=\left(
%\begin{array}{cc}
%0&1\\
%1&0\\
%\end{array}
%\right), \ \ \
%%%\quad
%\sigma_{2}=\left(
%\begin{array}{cc}
%0&-i\\
%i&0\\
%\end{array}
%\right), \ \  \
%%%\quad
%\sigma_{3}=\left(
%\begin{array}{cc}
%1&0\\
%0&-1\\
%\end{array}
%\right), \ \ \
%%%\quad
%\sigma_{0}=I=\left(
%\begin{array}{cc}
%1&0\\
%0&1\\
%\end{array}
%\right),
%$$
\[ \sigma_{1}=|0><1|+|1><0|=|0,1)+|1,0), \]
\[ \sigma_{2}=-i|0><1|+i|1><0|=-i(|0,1)-|1,0) ), \]
\[ \sigma_{3}=|0><0|-|1><1|=|0,0)-|1,1), \]
\[ \sigma_{0}=I=|0><0|+|1><1|=|0,0)+|1,1). \]
Let us use the formulas
\[ |0,0)=\frac{1}{2}(|\sigma_{0})+|\sigma_{3})) \ , \quad
|1,1)=\frac{1}{2}(|\sigma_{0})-|\sigma_{3})), \]
\[ |0,1)=\frac{1}{2}(|\sigma_{1})+i|\sigma_{2})) \ , \quad
|1,0)=\frac{1}{2}(|\sigma_{1})-i|\sigma_{2})). \]
It allows to rewrite operator basis
\[ |k,l)=|k_{1},l_{1})\otimes |k_{2},l_{2}) \otimes ...
\otimes |k_{n},l_{n}) \]
by complete basis operators
\[ |\sigma_{\mu})=|\sigma_{\mu_1} \otimes \sigma_{\mu_2} \otimes ...
\otimes \sigma_{\mu_n}), \]
where $\mu_i=2k_{i}+l_{i}$, i.e. $\mu_{i} \in \{0,1,2,3\}$ and $i=1,...,n$.
The basis $|\sigma_{\mu})$ is orthogonal
$(\sigma_{\mu}|\sigma_{\mu '})=2^{n} \delta_{\mu \mu '}$
and $|\sigma_{\mu})$ is complete operator basis
\[ \frac{1}{2^{n}} \sum^{N-1}_{\mu}
|\sigma_{\mu})(\sigma_{\mu}|=\hat I. \]
For an arbitrary element $|A)$ of $\overline{\cal H}^{(n)}$ we have
Pauli representation by
\[ |A)=\frac{1}{2^{n}}\sum^{N-1}_{\mu=0}
|\sigma_{\mu})A_{\mu} \]
with the complex coefficients
$A_{\mu}=(\sigma_{\mu}|A)=Tr( \sigma_{\mu} A )$. 
%%%%%%%%%%%%%%%%%%%%%%%%%%%%%%%%%%%%%%%%%%%%%%%%%%%%%%%%%%%%%%%%%%%%%%%%%%

\section{Appendix II. }

%%%%%%%%%%%%%%%%%%%%%%%%%%%%%%%%%%%%%%%%%%%%%%%%%%%%%%%%%%%%%%%%%%%%%%%%%%

Let us consider some elements of classical four-valued logic.
For the concept of many-valued logic see \cite{Re,RT,Ya2}.

A classical four-valued logic gate is called a function
$g(x_{1},...,x_{n})$ if 
all $x_{i} \in \{0,1,2,3\}$, where $i=1,...,n$, and
$g(x_{1},...,x_{n}) \in \{0,1,2,3\}$.

It is known that
the number of all classical logic gates with n-arguments
$x_{1},...,x_{n}$ is equal to
$4^{4^{n}}$.
The number of  classical logic gates $g(x)$ with single argument
is equal to $4^{4^{1}}=256$.

\vskip 5mm
\begin{tabular}{|c|c|c|c|c|c|c|c|c|}
\hline
\multicolumn{9}{|c|}{Single argument classical gates}\\
\hline
%%%%%%%%%%%%% [] is \Box
x& $\sim x$ &$\Box x$&$\diamondsuit x$&$\overline{x}$&$I_{0}$&$I_{1}$&$I_{2}$&$I_{3}$\\ \hline
0& 3        &   0    &      0         &      1       &  3    &   0   &   0   & 0   \\
1& 2        &   0    &      3         &      2       &  0    &   3   &   0   & 0    \\
2& 1        &   0    &      3         &      3       &  0    &   0   &   3   & 0   \\
3& 0        &   3    &      3         &      0       &  0    &   0   &   0   & 3    \\ \hline
\end{tabular}
\vskip 3mm

The number of  classical logic gates $g(x_{1},x_{2})$ with
two-arguments is equal to
$4^{4^{2}}=4^{16}=42949677296$. 

Let us define some elementary classical 
4-valued logic gates by formulas:\\
Luckasiewicz negation: \  $\sim x=3-x$.\\
Cyclic shift: \ $\overline{x}=x+1(mod4)$.\\
Functions $I_{i}(x)$, where $i=0,...,3$, such that
$I_{i}(x)=3$ if $x=i$ and $I_{i}(x)=0$ if $x\not=i$.\\
Generalized conjunction: \ $x_{1} \land x_{2}=min(x_{1},x_{2})$.\\
Generalized disjunction: \ $x_{1} \lor x_{2}=max(x_{1},x_{2})$.\\
Generalized Sheffer-Webb function: \
$$V_{4}(x_{1},x_{2})=max(x_{1},x_{2})+1(mod 4).$$

Commutative law, associative law and distributive law for the
generalized conjunction and disjunction are satisfied:\\
Commutative law:
\[ x_{1} \land x_{2}=x_{2} \land x_{1} \ , \quad
x_{1} \lor x_{2}=x_{2} \lor x_{1}. \]
Associative law:
\[ (x_{1} \lor x_{2}) \lor x_{3}=
x_{1} \lor (x_{2} \lor x_{3}). \]
\[ (x_{1} \land x_{2}) \land x_{3}=
x_{1} \land (x_{2} \land x_{3}). \]
Distributive law:
\[ x_{1}\lor (x_{2} \land x_{3})=
(x_{1} \lor x_{2}) \land ( x_{1} \lor x_{3}). \]
\[ x_{1}\land (x_{2} \lor x_{3})=
(x_{1} \land x_{2}) \lor ( x_{1} \land x_{3}). \]
Note that the Luckasiewicz negation is satisfied
\[ \sim(\sim x)=x \ , \quad
\sim(x_{1} \land x_{2})=(\sim x_{1}) \lor (\sim x_{2}). \]
The shift $\overline{x}$ for $x$ is not satisfied usual negation rules:
\[ \overline{\overline{x}} \not=x \ , \quad \overline{x_{1}\land x_{2}}
\not=\overline{x_{1}} \lor \overline{x_{2}}. \] The analog of
disjunction normal form of the n-arguments 4-valued logic gate is
\[ g(x_{1},...,x_{n})=\bigvee_{(k_{1},...,k_{n})} \ I_{k_{1}}(x_{1})
\land ... \land I_{k_{n}} \land g(k_{1},..,k_{n}). \]

It is known universal sets of universal classical gates of four-valued
logic:\\
{\it
The set $\{0, 1, 2, 3, I_{0}, I_{1}, I_{2}, I_{3},
x_{1} \land x_{2}, x_{1} \lor x_{2}\}$ is universal.\\
The set $\{\overline{x}, x_{1} \lor x_{2}\}$ is universal.\\
The gate $V_{4}(x_{1}, x_{2})$ is universal.}\\
This theorem is proved in \cite{Ya2}.

%%%%%%%%%%%%%%%%%%%%%%%%%%%%%%%%%%%%%%%%%%%%%%%%%%%%%%%%%%%%%%%%%%%%%%%%%%
%%%%%%%%%%%%%%%%%%%%%%%%%%%%%%%%%%%%%%%%%%%%%%%%%%%%%%%%%%%%%%%%%%%%%%%%%%
%%\newpage

\end{multicols}{2}
%%\today

%%%%%%%%%%%%%%%%%%%%%%%%%%%%%%%%%%%%%%%%%%%%%%%%%%%%%%%%%%%%%%%%%%%%%%%%%%
%%%%%%%%%%%%%%%%%%%%%%%%%%%%%%%%%%%%%%%%%%%%%%%%%%%%%%%%%%%%%%%%%%%%%%%%%%
%%%%%%%%%%%%%%%%%%%%%%%%%%%%%%%%%%%%%%%%%%%%%%%%%%%%%%%%%%%%%%%%%%%%%%%%%%
%%%%%%%%%%%%%%%%%%%%%%%%%%%%%%%%%%%%%%%%%%%%%%%%%%%%%%%%%%%%%%%%%%%%%%%%%%
%%%%%%%%%%%%%%%%%%%%%%%%%%%%%%%%%%%%%%%%%%%%%%%%%%%%%%%%%%%%%%%%%%%%%%%%%%
%%%%%%%%%%%%%%%%%%%%%%%%%%%%%%%%%%%%%%%%%%%%%%%%%%%%%%%%%%%%%%%%%%%%%%%%%%
%%%%%%%%%%%%%%%%%%%%%%%%%%%%%%%%%%%%%%%%%%%%%%%%%%%%%%%%%%%%%%%%%%%%%%%%%%

\end{document}